\documentclass{aa}

\usepackage{graphicx}
\usepackage{txfonts}

\usepackage{xcolor}
\usepackage{siunitx}
\usepackage{multirow}
\usepackage{booktabs}

\usepackage[pdfpagelabels=false]{hyperref}
\hypersetup{colorlinks=true,linkcolor=blue,citecolor=blue,filecolor=blue,urlcolor=blue}

\usepackage[normalem]{ulem}

\DeclareSIUnit\au{au}
\DeclareSIUnit\Msun{\ensuremath{M_\odot}}
\DeclareSIUnit\year{yr}
\DeclareSIUnit\erg{erg}

\newcommand{\angstrom}{\text{\normalfont\AA}\xspace}
\newcommand{\degree}{\ensuremath{^{\circ}}\xspace}

\newcommand{\oiline}{$\mathrm{[OI]}\,6300\,\mathrm{\angstrom}$\xspace}
\newcommand{\Neline}{[NeII]\,12.81\,{\textmu m}\xspace}
\newcommand{\htline}{o-H$_2$\,2.12\,{\textmu m}\xspace}

\newcommand{\prodimo}{\textsc{ProDiMo}}
\newcommand{\mocassin}{\textsc{mocassin}}
\newcommand{\bfour}{\ensuremath{\beta}\textit{4}\xspace}
\newcommand{\bsix}{\ensuremath{\beta}\textit{6}\xspace}
\newcommand{\pe}{\textit{PE}\xspace}

\usepackage{xspace}
\newcommand{\nirv}{\textsc{nirvana-iii}\xspace}

\newcommand{\etaO}{\eta_{\rm O}}
\newcommand{\etaH}{\eta_{\rm H}}
\newcommand{\etaA}{\eta_{\rm A}}

\newcommand\cm{\,\rm cm}
\newcommand\s{\,\rm s}

\newcommand\betap{\beta_{\rm P}}
\newcommand{\LX}{L_{\rm X}}

\newcommand{\V}{\mathbf{v}}
\newcommand{\B}{\mathbf{B}}
\newcommand{\E}{\mathbf{\mathcal{E}}}

\begin{document} 

\graphicspath{{figures/}}

\title{From thermal to magnetic driving: spectral diagnostics of simulation-based magnetothermal disc wind models}

\author{
    Michael L. Weber\inst{1,2}\fnmsep\thanks{ \href{mailto:mweber@usm.lmu.de}{mweber@usm.lmu.de}}
    \and Eleftheria Sarafidou\inst{3, 4, 5}
    \and Christian Rab\inst{1,6}
    \and Oliver Gressel\inst{4}
    \and Barbara Ercolano\inst{1,2,6}
}

\institute{
    University Observatory, Faculty of Physics, Ludwig-Maximilians-Universität München, Scheinerstr. 1, 81679 Munich, Germany
    \and Excellence Cluster ORIGINS, Boltzmannstr. 2, 85748 Garching, Germany
    \and Astronomy Unit, Department of Physics and Astronomy, Queen Mary University of London, London E1 4NS, UK
    \and Leibniz-Institut für Astrophysik Potsdam (AIP), An der Sternwarte 16, 14482, Potsdam, Germany
    \and Institut f\"ur Physik und Astronomie, Universit\"at Potsdam,
     Karl-Liebknecht-Str. 24/25, 14476 Golm, Germany
    \and Max-Planck-Institut für extraterrestrische Physik, Giessenbachstrasse 1, 85748 Garching, Germany
}

\date{Received <date> / Accepted <date>}

\abstract{
 Disc winds driven by thermal and magnetic processes are thought to play a critical role in protoplanetary disc evolution. However, the relative contribution of each mechanism remains uncertain, particularly in light of their observational signatures.
}
{
We investigate whether spatially resolved emission and synthetic spectral line profiles can distinguish between thermally and magnetically driven winds in protoplanetary discs.
}
{
We modelled three disc wind scenarios with different levels of magnetisation: a relatively strongly magnetised wind ($\beta4$), a rather weakly magnetised wind ($\beta6$), and a purely photoevaporative wind (\pe). Using radiative transfer post-processing, we generated synthetic emission maps and line profiles for \oiline, \Neline, and \htline, and compared them with observational trends in the literature.
}
{
We find that the $\beta4$ model generally produces broader and more blueshifted low-velocity components across all tracers, consistent with compact emission regions and steep velocity gradients. The $\beta6$ and \pe models yield narrower profiles with smaller blueshifts, in better agreement with most observed narrow low-velocity components (NLVCs). We also find that some line profile diagnostics, such as the inclination at maximum centroid velocity, are not robust discriminants. However, the overall blueshift and full-width at half-maximum (FWHM) of the low-velocity components provide reliable constraints. The $\beta4$ model reproduces the most extreme blueshifted NLVCs in observations, while most observed winds are more consistent with the $\beta6$ and \pe models.
}
{
Our findings reinforce previous conclusions that most observed NLVCs are compatible with weakly magnetised or purely photoevaporative flows. The combination of line kinematics and emission morphology offers meaningful constraints on wind-driving physics, and synthetic line modelling remains a powerful tool for probing disc wind mechanisms.
}

\authorrunning{Weber, M. et al.}
\titlerunning{Spectral diagnostics of simulation-based magnetothermal disc wind models}
\maketitle

\section{Introduction}

Disc winds are a fundamental component of protoplanetary disc evolution, with significant implications for both disc dispersal and planet formation. Over the past two decades, theoretical and observational efforts have increasingly focused on understanding the physical mechanisms that launch these winds and their observable consequences. In particular, forbidden emission lines such as \oiline and \Neline have become powerful tracers of disc winds, providing constraints on their thermal and kinematic properties \citep[e.g.][]{Ercolano2017,pascucci2020,banzatti2019}.

From a theoretical standpoint, two primary mechanisms have been proposed to drive disc winds: thermal photoevaporation and magnetically driven winds. Thermal models, particularly those driven by X-ray irradiation, have been extensively developed in the context of viscously evolving discs \citep[e.g.][]{owen2010,picogna2019,ercolano_picogna2022}. These models predict mass-loss rates and emission features that are broadly consistent with observations of transition discs and low-velocity forbidden line components \citep{ercolano2010,ercolano2016}. More recently, the importance of magnetised winds has gained traction, both in global MHD simulations \citep[][]{gressel2015,bai2016,bai2017,bethune2017} and semi-analytic models \citep[e.g.][]{suzuki2016,baigoodman2016,Lesur2021,tabone2022,lesur2023,kadam2025}. Initial efforts to compare synthetic line diagnostics for models containing both wind components have been attempted by bootstrapping analytical MHD wind solutions onto photoevaporative models \citep{weber2020}. At the same time, dynamical simulations combining non-ideal MHD evolution and detailed thermochemistry are rare \citep[but see][]{wang2019,gressel2020,sarafidou2024,sarafidou2025}. In the current paper, we aim to directly use outflow configurations obtained from non-ideal MHD simulations \citep[very similar to the ones presented in][]{sarafidou2024}, and derive a set of synthetic spectral diagnostics.

From an observational point of view, distinguishing between thermal and magnetic driving remains a key challenge. In recent years, several works have attempted to constrain wind-launching mechanisms through detailed modelling of emission lines. Notably, \citet{weber2020} and \citet{picogna2019} performed radiative transfer calculations of X-ray-driven photoevaporative winds and compared synthetic line profiles with observations. \citet{rab2022} further explored emission line diagnostics for a range of atomic and molecular species using thermo-chemical modelling. The observational community has provided increasingly rich datasets, particularly through high-resolution spectroscopy of T Tauri stars \citep[e.g.][]{pascucci2020,banzatti2019,fang2018,gangi2020,whelan2021}; however, even with spatially resolved observations, it remains challenging to distinguish the two different wind driving mechanisms \citep{fang2023,rab2023}.

These studies have revealed the presence of multiple kinematic components in forbidden lines, including broad and narrow low-velocity components (BLVCs and NLVCs), whose origins are still debated. While photoevaporative models can account for the NLVCs in many cases, BLVCs are generally associated with more compact inner disc regions and may require additional mechanisms such as magnetothermal launching or magnetospheric accretion funnels \citep[see, e.g.][]{takasao2018,takasao2022,zhu2024}.

Despite these advances, a systematic comparison of spectral line diagnostics across a sequence of magnetisation levels within self-consistent wind models has been lacking. In this paper, we address this gap by performing detailed post-processing of magnetothermal disc wind simulations with varying magnetic field strengths. We compute synthetic line profiles and emission maps for key tracers ([OI], [NeII], o-H$_2$) and evaluate their diagnostic potential for distinguishing wind-driving mechanisms. 

This work builds on the disc wind models of \citet{sarafidou2024} and radiative transfer techniques developed in our previous work \citep{weber2020,rab2022}. It complements the theoretical framework laid out in the \textit{Protostars and Planets VII} review by \citet{lesur2023} and the observational synthesis presented by \citet{pascucci2023}. Our goal is to provide a systematic framework for interpreting forbidden atomic and molecular hydrogen line observations in the context of competing wind-launching mechanisms.

 In Section 2, we describe the wind models and radiative transfer methodology used to generate the synthetic observables. Section 3 presents the results, including the structure of the wind and line emission regions. In Section 4, we discuss the implications of our findings, and in Section 5, we summarise our conclusions.

\section{Methods}
\subsection{Magnetothermal wind models}

The disc wind models performed here are based on the work of \citet{sarafidou2024}. We briefly outline the main features of these models; for a more detailed description, refer to that paper.

The simulations were carried out using the fluid \nirv code, which employs a standard second-order accurate finite volume scheme \citep{ziegler2004, ziegler2016}. They are 2D axisymmetric, using a spherical-polar coordinate system $(r, \theta, \phi)$ corresponding to radius, co-latitude, and azimuth. The computational domain spans $r \in (0.45, 16)\,\mathrm{au}$ and $\theta \in (0, \frac{1}{2}\pi)$, with a standard grid resolution of $N_r \times N_{\theta} = 456 \times 288$, and the state variables in the grid cells are updated using the intercell fluxes calculated by the Harten-Lax-van Leer Riemann solver \citep{harten1983}.

\subsubsection{Equations of motion}

The equations of motion for mass, momentum, and total energy densities are solved in conservation form, together with the induction equation for the magnetic field:

\begin{eqnarray}
  \partial_t\rho+\nabla\!\cdot(\rho\V) & = & 0\,,
  \nonumber\\
  \partial_t(\rho\V)+\nabla\!\cdot\big[\,\rho\V\V
    +P \mathbf{I}-\B\B\,\big] & = & \!-\rho\nabla\Phi\,,
  \nonumber\\
  \partial_t e + \nabla\!\cdot
     \big[\, (e+P)\V - (\V\cdot\B)\B\, \big]
     & = & \!-\rho(\nabla\Phi)\!\cdot\V + \nabla\!\cdot\!\mathcal{S} \,,
     \nonumber\\
  \partial_t \B - \nabla\times(\V\times\B) & = & - \nabla\times\E\,.
     \nonumber
\end{eqnarray}

Here, the total pressure $P$ is the sum of gas and magnetic pressures, and $\mathcal{S} \equiv \E\times\B$ denotes the Poynting flux. The electromotive force $\E$ incorporates the three non-ideal MHD effects, given by:

\begin{equation}
  \E \equiv \etaO\mathbf{J} \,+\, \etaH\mathbf{J}\times\hat{\mathbf{B}} \,+\, \etaA\mathbf{J}\times\hat{\mathbf{B}}\times\hat{\mathbf{B}}.
\end{equation}

where $\etaO$, $\etaA$, and $\etaH$ are the diffusion coefficients for Ohmic resistivity, ambipolar diffusion, and the Hall effect, respectively:

\begin{subequations}\label{eq:diffusivities}
  \begin{gather}
    \etaO = \frac{c^2\gamma_e m_e \rho}{4\pi\rho^2 n_e}\,,\qquad
    \etaA = \frac{B^2}{4\pi\gamma_i \rho\rho_i}\,,\qquad
    \etaH  = \frac{cB}{4\pi e n_e} \tag{\theequation a-c}
  \end{gather}
\end{subequations}

In these expressions, $n_e$, $e$, and $m_e$ represent the electron number density, charge, and mass respectively; $\rho$ is the density of neutrals; $\rho_i$ the ion density; and $\gamma = \langle\sigma u\rangle/(m_n+m_i)$ is the ion–neutral drift coefficient. The quantity $\langle\sigma u\rangle$ is the ion–neutral collision rate, with $\sigma$ denoting the conductivity, which for a proton–electron plasma is given by $\sigma = n_e e^2/(m_e f_c)$, where $f_c$ is the collision frequency. Finally, $\mathbf{J} = \nabla\times\B$ is the electric current density \citep[see, e.g.,][]{wardle_ng1999}.

Previous studies \citep[e.g., ][]{Lesur2021,sarafidou2024} have shown that the wind mass-loss rate is largely insensitive to the detailed microphysics of the disc. Since this work focuses on the wind properties, we therefore neglect the Hall effect by setting $\etaH = 0$, without expecting any significant impact on our results.

\subsubsection{Initial disc model}\label{sec:disk_model}

The simulations are initialised with the following equilibrium structure \citep[see][]{nelson2013}:  we adopt a power-law relation between the locally isothermal temperature $T$ and the cylindrical radius $R$,

\begin{equation}\label{eq: eqTemp}
 T(R) = T_0\left(\frac{R}{R_0}\right)^q\,,
\end{equation}

and a similar power-law relation for the midplane density:

\begin{equation}\label{eq: eqrho_mid}
 \rho_{\rm mid}(R) = \rho_0\left(\frac{R}{R_0}\right)^p\,.
\end{equation}

We adopt a temperature slope of $q = -0.5$, which corresponds to a mildly flared disc height as a function of radius, and a density slope of $p = -2.25$, consistent with typical observed PPD profiles \citep[e.g.,][]{manara2023}. The resulting equilibrium solutions are:

\begin{eqnarray}
  \rho(\mathbf{r}) & = & \rho_{\rm mid}(R) \exp{\left[\frac{GM_\star}{c_s^2} \left( \frac{1}{r}-\frac{1}{R}\right)\right]}\,
  \label{eq: dens_eq_sol}\\
  \Omega(\mathbf{r}) & = & \Omega_{\rm K}(R)\, \left[\,(p+q)\left(\frac{H}{R}\right)^2+(1+q)-\frac{qR}{r}\,\right]^{1/2}\,
  \label{eq: Omega_sol}
\end{eqnarray}

where $\Omega_{\rm K}(R) \equiv (GM_\star R^{-3})^{1/2}$ is the Keplerian angular velocity, $c_s^2(R) = c_{s0}^2(R/R_0)^q$ is the squared sound speed, and $H(R_0) = c_{s0}/\Omega_K(R_0) = 0.055\,R_0$ sets the pressure scale height of the initial disc. The initial density profile is scaled to $\Sigma_0 = 200\,{\rm g\,cm^{-2}}$ at $R_0 = 1 \,\mathrm{au}$, with an aspect ratio $H(R) = 0.055(R/R_0)^{1/4}$. We assume a central stellar mass $M_\star = 0.7M_{\odot}$, an adiabatic index $\gamma = 5/3$ for an ideal atomic gas, and a mean molecular weight $\mu \approx 1.37$, representing a typical light-element composition.

The initial magnetic field strength is controlled via the dimensionless plasma parameter $\betap$, which measures the ratio of thermal to magnetic pressure and serves as a standard indicator of the relative importance of magnetic effects:
\begin{equation}\label{eq:plasma_beta}
\betap \equiv \frac{2\mu_0\,\rho c_s^2}{B^2}\,.
\end{equation}
The code takes the midplane value $\beta_{\text p0}$ as an input parameter, from which $B_0$ is determined. The vertical magnetic field is then initialized according to
\begin{equation}
  \B(\mathbf{r}, t) = B_0\left(\frac{R}{R_0}\right)^{(p+q)/2}\hat{\mathbf{z}}\,,
\end{equation}
where the exponent reflects the radial dependence of the background density and temperature profiles.

We compute three models with identical inputs but different levels of magnetization, $\betap$. The most magnetized disc has $\betap = 10^{4}$ (model \bfour), followed by $\betap = 10^{6}$ (model \bsix), and finally a purely photoevaporative case with no magnetic field, denoted \pe.

\subsubsection{X-Ray heating prescription}\label{sec:xray_heating}

This work aims to investigate whether observational signatures can distinguish between magnetic and photoevaporative winds. For this we introduced a heating prescription for X-ray irradiation from the central star.

This formulation builds upon the work of \citet{owen2010} that was later refined by \citet{picogna2019}. The underlying parametrization assumes that in X-ray-irradiated regions, the gas temperature, $T_{\mathrm{gas}} = T_{\mathrm{X}}(\Sigma,\xi)$, depends only on the intervening column density, $\Sigma$, and the ionization parameter $\xi$. The rationale for this approach is detailed in \citet{ercolano_picogna2022} and references therein.

The ionization parameter is defined as  
\begin{equation}
\xi \equiv \frac{\LX}{n\,r^2}
\end{equation}  
where $n$ represents the local number density of the gas, and $\LX$ denotes the assumed X-ray luminosity. The function $T_{\mathrm{X}}(\Sigma,\xi)$ is tabulated using results from detailed thermal and ionization calculations performed with the \textsc{mocassin} code \citep{ercolano2003, ercolano2005, ercolano2008}.

We adopt an X-ray luminosity of $\LX = 2\times10^{30},{\rm erg,s^{-1}}$, using the tabulated parameters from Table~1 of \citet{picogna2019}.

\subsection{Synthetic observables} \label{sec:methods:profiles}
To produce synthetic observations of atomic species, specifically the \oiline and \Neline forbidden emission lines, we follow the approach by \citet{weber2020} and post process the models using the Monte Carlo radiative transfer code \mocassin{} \citep{ercolano2003, ercolano2005, ercolano2008}. Moreover, we use the thermo-chemical code \prodimo{} (PROtoplanetary DIsk MOdel\footnote{\url{https://prodimo.iwf.oeaw.ac.at} Version v3.0.0}, \citealt{woitke2009a,kamp2010,thi2011,woitke2016}), for the \htline molecular hydrogen line. The use of two different approaches is justified, as previous studies have shown that they yield consistent results for atomic species \citep{rab2022,rab2023}. In addition, \mocassin~provides spatially resolved emission maps, which are particularly useful for interpreting the synthetic observables.

In both approaches, we model the stellar spectrum using the EUV+X-ray spectrum presented in \citet{ercolano2009}, complemented by an additional component representing the UV emission produced by accretion onto the star.
We approximate this accretion-related UV component with a blackbody at $T = 12000$\,\si{K}, assigning it a luminosity derived from the mass accretion rate measured in the models.
The conversion follows the relation
\begin{equation}
	L_\mathrm{acc} = \left( 1 - \frac{R_*}{R_\mathrm{in}} \right) \frac{G M_* \dot{M}_\mathrm{acc}}{R_*} \approx 0.8 \frac{G M_* \dot{M}_\mathrm{acc}}{R_*},
\end{equation}
where we assume an inner truncation radius of $R_\mathrm{in} = 5\,R_*$ and adopt $R_* = 2\,R_\odot$.
The measured mass accretion rates and the resulting accretion luminosities are listed in \autoref{tab:lacc}.

Since our model is inviscid and accretion is entirely driven by magnetic stresses, the model does not exhibit a measurable accretion rate in the purely photoevaporative case (model \pe).
However, even in the absence of magnetic winds, protoplanetary discs are expected to accrete, potentially due to turbulence or other processes \citep[e.g.][]{manara2023}.
To enable a direct comparison with the weakly accreting \bsix model, we assume the same accretion luminosity for the \pe model, which effectively imposes an accretion rate in the \pe model that is not self-consistently generated.
While this does not significantly affect the wind structure, the resulting accretion luminosity has a strong impact on the observable signatures \citep{ercolano2016}.

We note that mass loss due to photoevaporation is independent of the accretion rate or luminosity, as it is primarily driven by X-ray irradiation.
The accretion luminosity component is only relevant for the excitation of wind-tracing observables.
It is therefore reasonable to assume that the overall wind structure remains unaffected when including the accretion component in the irradiating spectrum.
\begin{table}
	\centering
	\caption[Mass-loss and accretion rates and accretion luminosity]{Wind mass-loss rates and mass accretion rates measured in the models, as well as the accretion luminosities resulting from the mass accretion rate conversion.}
	\label{tab:lacc}
	\begin{tabular}{l|l|l|l}
		\toprule
		Model & $\dot{M}_\mathrm{wind}$ [\si{\Msun \per \year}] & $\dot{M}_\mathrm{acc}$ [\si{\Msun \per \year}] & $L_\mathrm{acc}$ [L$_\odot$] \\
		\midrule
		\bfour &  $4.20 \cdot 10^{-8}$  & $3\cdot10^{-8}$	& 0.26              \\
		\bsix  &  $1.90 \cdot 10^{-8}$  & $1\cdot10^{-9}$   & $8 \cdot 10^{-3}$ \\
		\pe     &  $1.67 \cdot 10^{-8}$  & --                & $8 \cdot 10^{-3}$ \\
		\bottomrule
	\end{tabular}
\end{table}

Before post-processing, we remap the models onto a Cartesian grid extending from 0 to \SI{16}{\au} in both directions on 600 quadratically spaced grid points.
Moreover, we exclude cells inside $r < 0.6$\,\si{au} to eliminate a spurious gas pile-up near the inner radial boundary, caused by the damping buffer zones.

Furthermore, the regions close to the polar axis can also exhibit an unphysical flow that arises from purely numerical effects \citep{gressel2020, sarafidou2024}.
To avoid contamination of the synthetic observables with emission from this region, we exclude cells that satisfy any of the following conditions: $R < 0.45$\,\si{\au}, $n < $\,\SI{1e5}{\per \centi \metre \cubed}, or ($n < $\,\SI{1e7}{\per \centi \metre \cubed} and {$v_z < 0$}), where $n$ is the local number density.
This filtering reliably removes the problematic regions without affecting the wind or the bound disc.

To compute spectral profiles of the \oiline and \Neline lines, we adopt the method of \citet{weber2020}, with the modification that dust extinction is neglected in this work. For the \htline line, we follow the approach of \citet{rab2022}.
In order to simulate realistic observational conditions, the resulting profiles are convolved with a Gaussian kernel of width $\mathrm{FWHM} = c / R_\mathrm{spec}$, where $c$ is the speed of light and $R_\mathrm{spec}$ is the spectral resolving power of the simulated instrument.
For the optical lines, we adopt a resolving power of $R_\mathrm{spec,opt} = 50000$, representative of most available datasets.
The infrared lines are degraded to $R_\mathrm{spec,IR} = 30000$.

\section{Results}

\subsection{Density and velocity structure}
\begin{figure*}
	\centering
	\includegraphics[width=.8\textwidth]{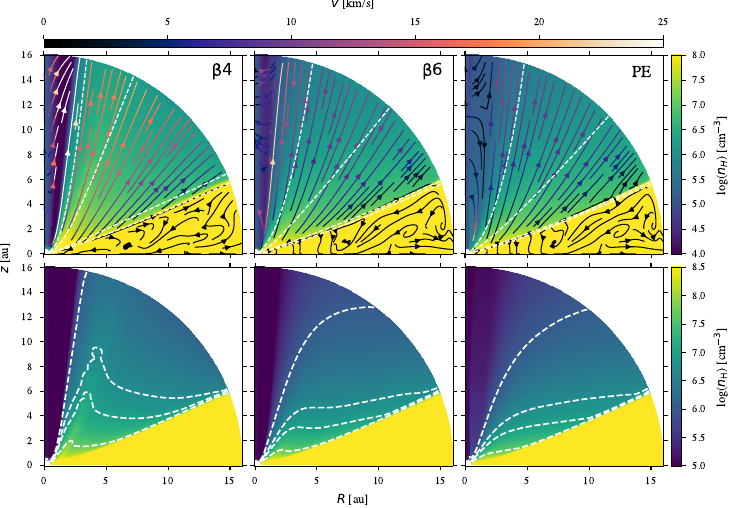}
	\caption[Number density maps of the magnetothermal models]{
		Number density maps for the three models \bfour, \bsix and \pe (columns from left to right). Top row: The streamlines represent the velocity structure, and their colour represents the speed. White dashed lines are column density contours for the values (from top to bottom) 10$^{20}$, 10$^{21}$, 10$^{22}, 2.5\cdot10^{22}$\,\si{\per \cm \squared}. Bottom row: White dashed lines are density contours for the values (from top to bottom) $10^6$, $5\cdot10^6$, $10^7$, and $5\cdot10^7$ \,\si{\per \centi \metre \cubed}.
	}
	\label{fig:models}
\end{figure*}
In \autoref{fig:models} we present the density and velocity structure of the wind models. All three models exhibit a sharp transition from high-density in the bound disc to low density at the base of the wind. Throughout this work, we will refer to this region as the disc-wind interface or the disc surface for simplicity.

The \bfour model exhibits a narrow, dense inner cone that emanates from a ``puffed-up'' inner disc at $R < 2$\,\si{\au}.
This cone significantly raises the column density to values \SI{\sim 1e21}{\per \cm \squared}, as seen from the dashed contour lines.
The wind reaches high velocities exceeding 25\,\si{\km \per \s} along the inner edge and experiences significant acceleration even at high heights ($\gtrsim 10$\,\si{\au}) above the disc surface. 

Model \bsix, where the magnetic field is weaker, does exhibit a similar ``puffed-up'' inner disc, but in contrast to model \bfour, there is no well-defined outflow cone rising from it. Overall, the wind density at small radii is lower and drops off much more quickly in the vertical direction. The lower density is also apparent in the column density contours, which reach much deeper into the wind in the \bsix model. 
The wind velocity is slower with speeds exceeding 15\,\si{\km \per \s} only along the innermost edge. Moreover, the wind is not or only weakly accelerated beyond a few \si{\au} above the disc surface.

Model \pe is very similar to \bsix. The biggest differences can be seen in the inner regions at $R \lesssim 3$\,\si{\au}, where the wind velocities remain below $\sim 10$\,\si{\km \per \s}, and the ``puffed-up'' disc is slightly less extended in the vertical direction. 
At larger radii, the density and velocity field of the wind remain consistent between both models, suggesting that the magnetic contribution to the wind-launching in the \bsix model is small and mainly confined to the inner regions. This is supported by the comparison of the wind mass-loss rates (see \autoref{tab:lacc}), which reveals only a $\sim 13$\,\si{\percent} higher rate in the \bsix model compared to the \pe model, while the mass-loss rate in the \bfour model is $\approx 2.5$ times higher.

\begin{figure}
	\centering
	\includegraphics[width=.4\textwidth]{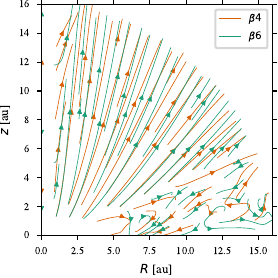}
	\caption[Streamline comparison]{
        Comparison of the velocity streamlines in the models \bfour (red) and \bsix (green).
	}
	\label{fig:streamlines}
\end{figure}
\autoref{fig:streamlines} shows a comparison of the streamlines between the models \bfour and \bsix.
In the wind-launching region close to the disc surface the behaviour depends on the radius. In the inner disc ($R \lesssim 5$\,\si{\au}), the wind is launched at a steeper angle when the hypersonic accretion in wind-emitting transitional discs is high (model \bfour), while the situation is reversed in the outer disc ($R \gtrsim 10$\,\si{\au}).
However, a few \si{\au} above the midplane the \bsix model has steeper outflow angles at all radii.

\subsection{Temperature and ionisation structure}
\begin{figure*}
	\centering
	\includegraphics[width=.8\textwidth]{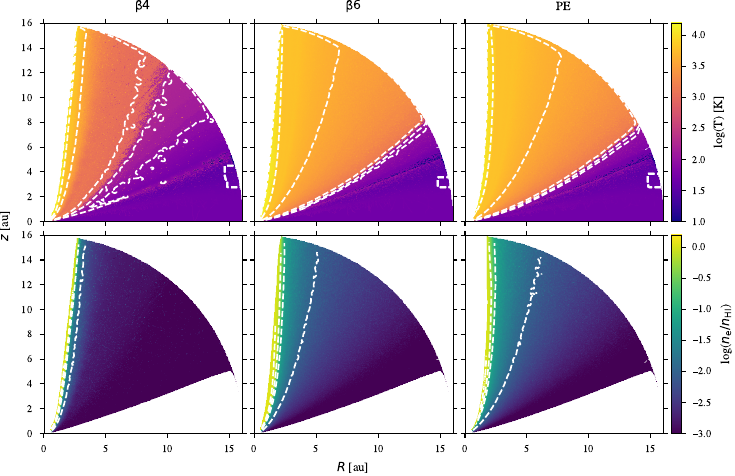}
	\caption[Temperature and ionisation structure]{
		  \mocassin~post-processing results. Top panel: Gas temperature. The white dashed lines are contours at 100, 200, 1000, 4000, and 8000\,\si{K}, smoothed with a Gaussian filter with $\sigma = 4$ to filter Monte-Carlo noise.
		Bottom panels: Ionisation fraction $n_\mathrm{e} / n_\mathrm{HI}$ and contours for values of $10^{-2}$, $10^{-1}$, 0.5, and 1, smoothed with a Gaussian filter with $\sigma = 2$.
	}
	\label{fig:thermal-structure}
\end{figure*}
\autoref{fig:thermal-structure} presents the thermal and ionisation structure of the models, obtained after post-processing them with \textsc{mocassin}.
In the \bfour model the majority of the wind region is much cooler than in the models with lower magnetisation.
Temperatures above 4000\,\si{\K} are reached only in a narrow band along the inner edge of the wind with a radial width $< 2$\,\si{\au}.
In contrast, the 4000\,\si{\K} contours of the \bsix and \pe models reach far deeper into the wind and are comparable in extent to the 1000\,\si{\K} contour in the \bfour model.
Between the two, the \pe model maintains slightly higher temperatures than the \bsix model throughout much of the wind.

The bottom panels of \autoref{fig:thermal-structure} show the ionisation fraction in the wind, which has a similar behaviour.
In all models there is a narrow, hot ($T > 8000$\,\si{K}) ionisation front along the inner edge of the wind but in model \bfour the fraction drops to below 1\,\% within less than 2\,\si{\au} in the radial direction, whereas the other models are lowly ionised ($\lesssim 10$\,\%) for several \si{\au} into the wind.

\subsection{Line luminosities \& emission regions}
\begin{figure}
	\centering
	\includegraphics[width=.4\textwidth]{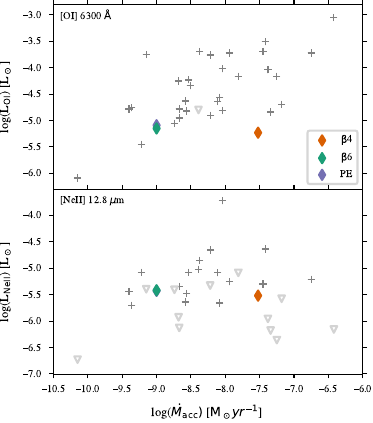}
	\caption[$L_\mathrm{line} - \dot{M}_\mathrm{acc}$ correlation]{
		\oiline and \Neline luminosities against the accretion rate in the models compared with the observations (grey symbols) by \citet{pascucci2020}. Triangles indicate upper limits.
	}
	\label{fig:L-Macc}
\end{figure}
Multiple studies have shown that \oiline luminosity positively correlates with accretion luminosity and therefore with the mass accretion rate \citep[e.g.][]{simon2016, banzatti2019}.
This correlation has been successfully reproduced in models of photoevaporating discs and can be explained by a positive relationship between the accretion luminosity and the size of the hot ($T \gtrsim 4000$\,\si{\K}) neutral wind region, which is predominantly heated by the EUV component of the accretion luminosity \citep{ercolano2016, weber2020}.
In \autoref{fig:L-Macc} we present the \oiline and \Neline luminosities of our models as a function of the mass accretion rate.
For comparison, we include the observations from \citet{pascucci2020}. 
Contrary to expectations, the \oiline luminosity remains approximately constant among all three models and even slightly decreases in the \bfour model.
While the \bsix and \pe models lie within the observed range, the \oiline luminosity in the \bfour model falls below the values typically observed in objects with similar accretion rates.

For the \Neline line, observations show no clear evidence of a correlation with accretion luminosity.
Theoretically, no such correlation is expected either, as the accretion luminosity - modeled here as a blackbody at 12000\,\si{\K} - is not energetic enough to ionise neon.
As shown in the bottom panel of \autoref{fig:L-Macc}, the \Neline luminosities in all models are similar and lie within the observed range.

The exact values of the luminosities measured in our models are listed in \autoref{tab:fluxes}.
\begin{table}
	\centering
	\caption[Line luminosities]{Line luminosities given as $\log_{10}(L_\mathrm{line} / L_\odot$)}
	\label{tab:fluxes}
    \begin{tabular}{l|l|l|l}
        \toprule
        Line & \bfour & \bsix & \pe \\
        \midrule
        \oiline & -5.23 & -5.15 & -5.09 \\
        \Neline & -5.52 & -5.42 & -5.43 \\
        \htline  & -6.72 & -7.06 & -7.08 \\
\bottomrule
    \end{tabular}
\end{table}

\begin{figure*}[!tbp]
	\centering
	\includegraphics[width=.8\textwidth]{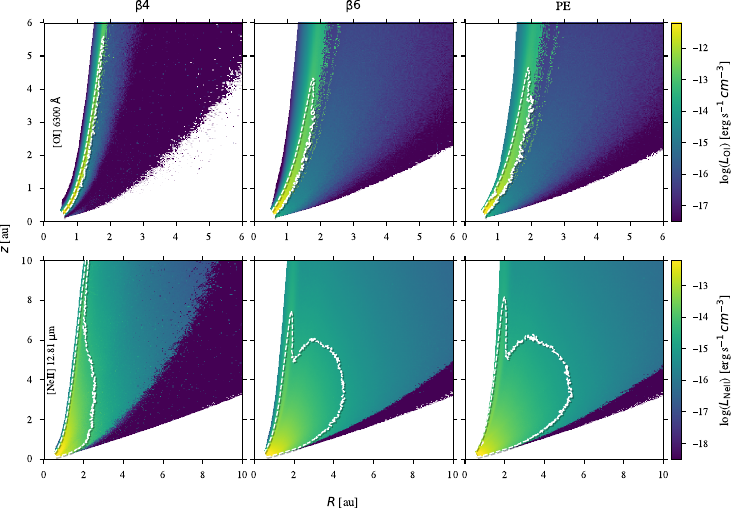}
	\caption[Emissivity maps]{
		Emissivity maps of the \oiline and \Neline lines. The white dashed lines are contours enclosing the region where 80\% of the total line flux is emitted in a 3D axisymmetric disc.
	}
	\label{fig:emission-maps}
\end{figure*}
In \autoref{fig:emission-maps}, we present maps of the emissivities of the \oiline and \Neline lines, overlaid with contours that enclose the volume from which 80\% of the total line luminosity originates.
The \oiline line traces a vertically extended layer near the EUV ionisation front.
In model \bfour, this layer is narrower but more vertically extended than in the other models, reaching out to $R \sim 1.75$\,\si{\au} and rising to $z \sim 5.6$\,\si{\au} at its highest point.
The \bsix\ (and \pe) models extend to $R \sim 1.75$ (2)\,\si{\au} at $z \sim 3$ (3.5)\,\si{\au}, with a radially narrower tip reaching up to $z \sim 4.3$ (4.5)\,\si{\au}.

The \Neline line behaves in a similar way but extends much farther into the wind, with 80\% contours reaching out to $R \sim 2.5$, 4.4, and 5.2\,\si{\au} in models \bfour, \bsix, and \pe, respectively.
In all models, the contours reach their maximum radial extent at a height of approximately $z \sim 3$\,\si{\au}.
In addition to this radially extended component, all models also feature a narrow, vertically extended emission region along the inner edge of the wind, reaching heights of $z \sim 14$, 7.5, and 8\,\si{\au}, respectively.
The reduced spatial extent of the \Neline emission in the \bfour model compared to the other models suggests that a significant fraction of the soft X-ray photons responsible for ionising neon is attenuated before reaching the outer wind.
This interpretation is supported by the strong inner wind column densities in the \bfour model, which efficiently absorb soft X-rays and limit the size of the ionised region.

It is important to note that in all models, the line emissivity exhibits a steep spatial gradient.
As a result, the 80\% luminosity contours primarily highlight the general extent of the emission region but do not capture the underlying variation in local emissivity. This is particularly relevant when comparing different lines, as their emission may be concentrated in distinct subregions within those contours. Therefore, flux ratios between lines with differing spatial distributions should be interpreted with care.

\begin{figure*}[!tbp]
	\centering
	\includegraphics[width=.8\textwidth]{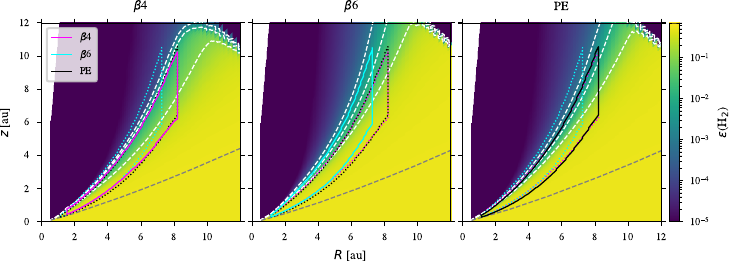}
	\caption[H$_2$ emission regions]{
		H$_2$ abundance obtained by processing the models with \prodimo{}. The white dashed lines are contours for values of (from top to bottom) 10$^{-3}$, 10$^{-2}$, and 10$^{-1}$. The grey dashed lines represent the $2.5\cdot10^{22}$\,\si{\per \cm \squared} radial column number density contour. The solid lines indicate where the \htline emission reaches 15\% and 85\% in the radial (integrated inside-out) and vertical (integrated from top to bottom) directions. For better comparison, each panel contains also the emission regions of the other models, shown as dotted lines.
	}
	\label{fig:h2-abundance}
\end{figure*}
In \autoref{fig:h2-abundance}, we present maps of the H$_2$ abundance and \htline emission regions obtained by processing the models with \prodimo{}.
Due to technical differences in the post-processing for molecular species, the emission regions in this figure are illustrated differently, with the contours indicating where the emission reaches 15\% and 85\% of its total value integrated from inside-out in the radial direction and from top to bottom in the vertical direction, respectively. 
The H$_2$ abundance in the \bfour{} model shows a relatively confined transition region between low ($< 10^{-5}$) and higher ($> 10^{-2}$) abundances compared to the \bsix and \pe models, where the transition appears more extended and gradual.
However, an interesting feature is that, despite the overall similarity between the \bsix and \pe models, H$_2$ in the \bsix model appears to be abundant and the \htline emitted higher up and closer in compared to the \pe model.
In contrast, the \bfour model shows an \htline emission region that aligns remarkably well with the \pe model, despite having a very different overall structure.
This behavior is the result of a subtle interplay between photodissociation, density distribution, and inner disc geometry.
In the \bsix model, a slightly denser inner wind cone and a more inflated inner disc act to shield part of the EUV radiation, protecting the H$_2$ from photodissociation.
As a result, the H$_2$ forms higher up and closer to the star.
In the \pe model, this EUV shielding is weaker, allowing H$_2$ to form deeper and further out in the flow.
In the \bfour model, although the screening is even stronger, the higher accretion rate leads to a correspondingly higher EUV luminosity, which compensates for the attenuation and results in an \htline emission region that closely aligns with that of the \pe model.

\subsection{Spectral line profiles}

\subsubsection{Profiles}
\begin{figure*}
	\centering
	\includegraphics[width=.8\textwidth]{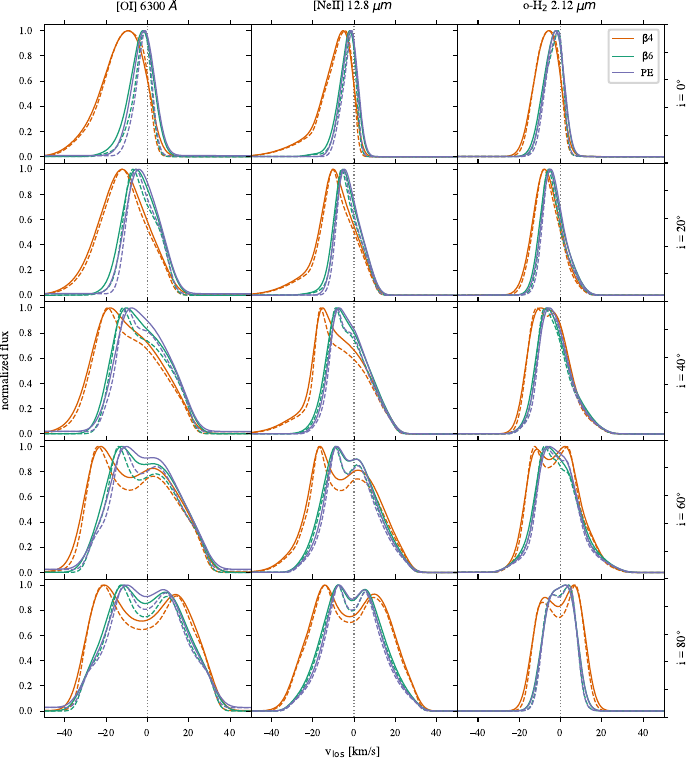}
	\caption[Line profiles]{
		Synthetic spectral line profiles observed at different inclinations. Solid lines represent the profiles degraded to a resolving power $R_\mathrm{spec, opt} = 50000$ for the \oiline line and $R_\mathrm{spec, IR} = 30000$ for the IR lines. Dashed lines represent profiles simulated with twice the resolving power (2$R$).
	}
	\label{fig:profiles}
\end{figure*}
In \autoref{fig:profiles}, we present synthetic spectral line profiles of \oiline, \Neline, and \htline, computed from the models at five inclinations ranging from $i = 0\degree$ (face-on) to $80\degree$ (nearly edge-on).
The lines have been normalised to their respective peak fluxes to facilitate a comparison of their shapes. The profiles have been convolved to a spectral resolution of $R = 50000$ for the \oiline line, and $R = 30000$ for the remaining lines to simulate realistic observational conditions. For comparison, profiles with twice this resolution are also included in the figure. Since the lower-resolution spectra capture all the most prominent features, the following arguments and conclusions are also valid for high-resolution spectra.

The \bfour model produces the broadest and most blueshifted profiles of all models at all inclinations, reflecting the higher wind speed in the radially confined emission regions along the inner edge of the magnetically dominated wind.
This difference becomes smaller, as the emission region of the line extends to larger radii, i.e. it is strongest for the \oiline line and weakest for the \htline line.
The \bsix and \pe profiles are nearly indistinguishable across all lines and inclinations, exhibiting similar overall shapes and extents.
However, the \bsix profiles tend to be slightly more blueshifted on the blue side, indicating slightly higher outflow speeds in the lowly magnetised model.

All line profiles follow the same general trend:
At low inclinations ($i \leq 20\degree$), the profiles are single-peaked.
The \bsix and \pe profiles are blueshifted by a few km/s, while the \bfour profiles exhibit larger blueshifts. 
At intermediate inclinations, the profiles become increasingly blueshifted and asymmetric with a noticeable shoulder or secondary peak on the red side.
While the \bsix and \pe profiles are only slightly blueshifted and narrow, the \bfour profiles are significantly broader and more blueshifted with an extended blueshifted tail, due to the large velocity gradient in the emission region.
In the \bsix and \pe models, the line wings broaden significantly on both the blue- and redshifted sides, consistent with Keplerian rotation being the dominant source of line broadening. In contrast, the \bfour profiles show much less broadening on the blueshifted side, while the red wing extends noticeably.
This suggests that in the \bfour model, the broadening on the blueshifted side is primarily driven by the wind’s velocity gradient rather than by Keplerian rotation.
At high inclinations ($i \geq 60\degree$), the profiles become more symmetric again and develop a double-peaked structure, where the peak on the blueshifted side is higher, with an exception of the \htline profiles. The \htline profiles likely behave differently due to their larger radial emission regions, and tend to show also self-absorption features for ($i \geq 60\degree$), for further details see \citet{rab2022}.

\subsubsection{Multi-Gaussian Fits}
\begin{figure}
	\centering
	\includegraphics[width=.48\textwidth]{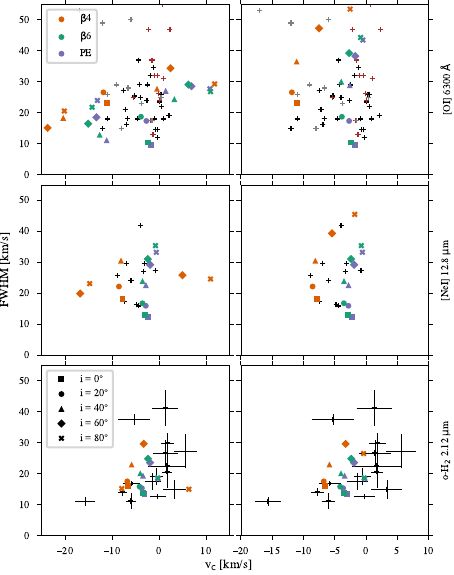}
	\caption[Overview of line centroids and FWHMs]{
		Overview of the line centroid and FWHMs of the Gaussian components that fit the profiles.
		The panels on the left show the results of the multi-Gaussian decomposition.
		The panels on the right show the properties when the profile is fit with a singular Gaussian component.
		For the \oiline line, black plus markers represent the NLVCs reported by \citet{banzatti2019, fang2018}, grey and brown plus markers represent SCJ and SC components by \citet{banzatti2019}, respectively.
		Black plus markers in the \Neline line panels are the observed components reported by \citet{pascucci2020}.
        Black plus markers with error bars in the \htline line panels are the observed components reported by \citet{gangi2020}.
    }
	\label{fig:vc-fwhm}
\end{figure}
To facilitate the analysis and interpretation of a large number of observational samples, observed profiles are usually decomposed into multiple Gaussian components.
In order to compare our synthetic profiles to the observational samples by \citet{banzatti2019, fang2018, pascucci2020}, and \citet{gangi2020},  we perform a similar decomposition by applying the multi-Gaussian fitting procedure described in \autoref{sec:methods:profiles}.
The resulting centroid-velocities and FWHMs of the individual Gaussian components are shown in the left panels of \autoref{fig:vc-fwhm}.
Since our synthetic profiles frequently exhibit double-peaked structures that are more prominent than typically seen in observations, the multi-Gaussian decomposition often yields two components, one significantly blueshifted and the other redshifted.
This can complicate direct comparisons with observed samples, where such distinct velocity components are usually absent.
For this reason, we also present centroid velocities and FWHMs derived from single-Gaussian fits in the right panels of \autoref{fig:vc-fwhm}.
Our discussion will primarily focus on these single-component fits, as they provide a more direct and consistent basis for comparison with observations.
While fitting only a single Gaussian inevitably leads to a loss of detail, this simplification is justified.
In observed spectra, the NLVCs often overlap with other components, such as BLVCs, which are not fully reproduced by our models.
Consequently, the single-Gaussian approach offers a pragmatic compromise that reflects the limitations of both models and observational data.
For more detailed studies, it is essential to examine individual profiles and their full spectral shapes directly, rather than relying solely on Gaussian decomposition. Nevertheless, the decomposition remains a useful tool for summarizing and comparing overall trends across models.

With FWHMs below 40\,km/s, the majority of the synthetic components fall into the category of NLVCs.
The only exceptions are the \oiline profiles at $i \geq 60\degree$ in the \bfour model and at $i = 80\degree$ in the \bsix model, as well as the \Neline profile at $i = 80\degree$ in the \bfour model. 
These cases exceed the 40\,km/s threshold by up to 13\,km/s.
The general absence of BLVCs is expected, as the inner boundary of our models is at 0.6\,au, while BLVCs are believed to originate within $\sim$0.5\,au.
Overall, our set of models reproduces the full range of observed FWHMs well, with the exception of the broadest \htline components, which reach FWHMs of $\sim$40\,km/s in the observations.

As already evident from the shape of the line profiles, the \bfour model exhibits the strongest blueshifts, reflected in the centroid velocities.
While the \bsix and \pe models produce centroid blueshifts of only a few km/s across all three lines, the \bfour model reaches values exceeding 10\,km/s in \oiline, close to 10\,km/s in \Neline, and up to $\sim$7\,km/s in the \htline line.
In all cases, the centroid velocities in the \bfour model are consistent with the most blueshifted NLVCs observed.
Only one \oiline BLVC and one \htline NLVC show stronger blueshifts than those reproduced by \bfour.
This suggests that our \bfour model already represents an extreme case and stronger winds are rarely occurring in the regions traced by NLVCs.
The majority of observed \oiline and \Neline NLVCs have centroid velocities closer to zero, consistent with models \bsix and \pe.
For the \htline line, all models tend to produce centroids that are more blueshifted than observed at intermediate FWHMs (20–30\,km/s), although the observational sample in this range is limited.
A clear trend is visible across all tracers: higher magnetisation leads to more blueshifted centroids and broader components.

None of the models reproduce components centered on the redshifted side of the spectrum, which are commonly observed in the \oiline and \htline lines.
Moreover, a substantial fraction of observed \oiline components are centered near zero velocity with FWHMs $\lesssim30$\,km/s - features that are also not matched by our models.
However, many of them were classified as single components (SCs) by \citet{banzatti2019}, and are thought to trace more evolved discs with inner cavities.
Such components have been successfully reproduced in photoevaporation models of transition discs, where the receding side of the wind becomes visible through the cavity \citep{ercolano2010, picogna2019}.

\subsubsection{Correlation with inclination}
\begin{figure}
	\centering
	\includegraphics[width=.48\textwidth]{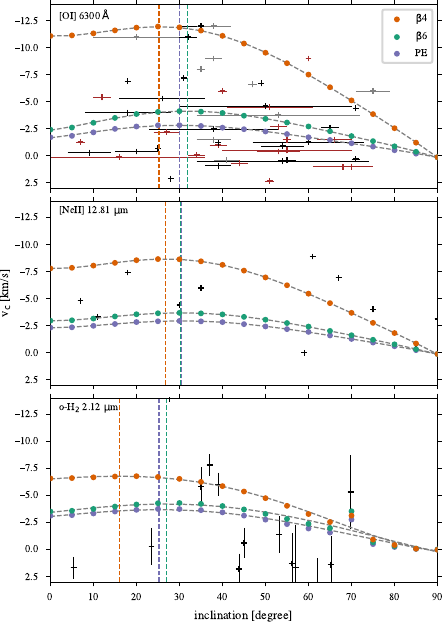}
	\caption[Centroid velocities as a function of inclination]{
		Centroid velocities against disc inclination. The dashed grey lines indicate a 5th-order polynomial fit to the centroid velocities. The dashed vertical lines indicate the inclination at the peak of these fits.
        Plus markers and errorbars as in \autoref{fig:vc-fwhm}.
	}
	\label{fig:vc-inc}
\end{figure}

In \autoref{fig:vc-inc}, we present the centroid velocities as a function of disc inclination.
The inclinations at which the centroid velocities reach their maximum are indicated by dashed vertical lines. 
These peak inclinations were identified by fitting a fifth-order polynomial to the model data. 
Differences in peak inclination between the models are generally modest. 
Even in the \htline line, where the \bfour\ and \pe\ models differ by nearly 10\degree, it would be difficult to determine the peak position observationally.
For instance, in the \bfour model, the centroid velocities at inclinations $i \leq 35\degree$ all lie within a narrow range of $[-6.7, -6.24]$\,km/s, resulting in a relatively flat trend that is difficult to distinguish observationally.

\begin{figure}
	\centering
	\includegraphics[width=.48\textwidth]{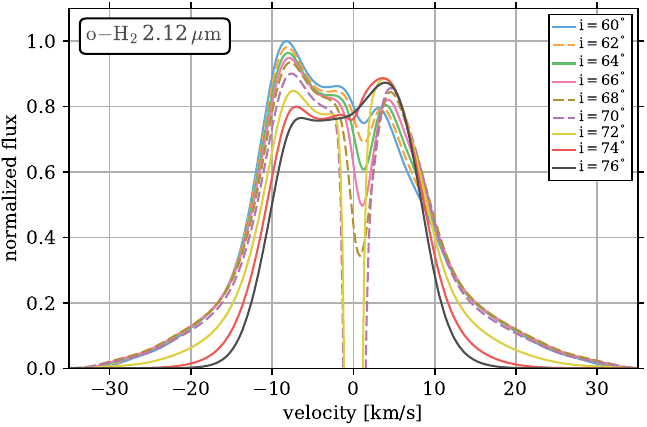}
	\caption[Self-absorption of \htline]{
		Line profiles of the \htline emission from the \bsix model at inclinations $60\degree \leq i \leq 76\degree$, illustrating the inclination-dependent self-absorption feature. The dashed lines are only used to improve clarity.
    }
	\label{fig:profile_h2_abs}
\end{figure}

An additional notable feature in the \htline line is an elevated centroid velocity at $i = 70\degree$, which appears in all models.
A closer inspection of the corresponding line profiles reveals a prominent absorption feature at this inclination.
The absorption introduces an asymmetry that is not well captured by a single-Gaussian fit, leading to a centroid that is more strongly blueshifted than it would be in the absence of the absorption \citep[see also][]{rab2022}.
This is illustrated in \autoref{fig:profile_h2_abs}, where the line profile of the \htline emission from the \bsix model is shown for multiple inclinations between 60\degree and 76\degree.

Comparable self-absorption features may also be present in observed profiles. 
For instance, \citet{pascucci2020} report a red-shifted peak in the \Neline profile of T Cha, resembling our predicted \htline profiles at high inclination ($i = 80\degree)$.
Another example is HN Tau, where one of the \htline profiles observed by \citet{gangi2020} hints at absorption, although no centroid velocity was reported for this profile.
However, in another epoch the feature is absent and the reported centroid velocity of -2\,km/s agrees well with our models.
These examples suggest that self-absorption signatures can occur in real systems, although their visibility depends strongly on inclination and other disc properties such as flaring.

When comparing both the \oiline and \htline lines with the observational sample, the centroid velocities from the \bfour model lie within 2\,km/s of the most blueshifted observations at all inclinations, further supporting our previous interpretation that the \bfour model represents a relatively extreme case.
In contrast, for the \Neline line, there are two observations at inclinations of $i \sim 60$–$65\degree$ where the observed centroids exceed the \bfour predictions by more than 2\,km/s.
However, the observed profiles of those two targets (V836 Tau, RY Lup) have relatively low signal-to-noise ratios, which makes an accurate measurement of the centroid velocity difficult \citep{pascucci2020}.

\begin{figure}
	\centering
	\includegraphics[width=.48\textwidth]{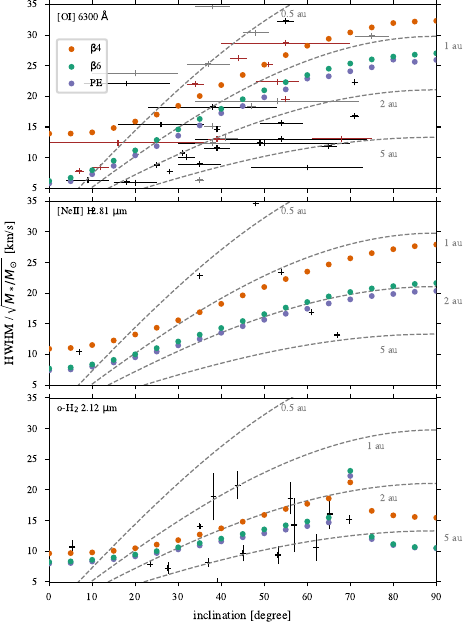}
	\caption[Normalized widths as a function of inclination]{
		Normalised component width as a function of disc inclination. The dashed grey lines indicate the projected velocity of a purely Keplerian disc at the indicated radius.
        Plus markers and errorbars as in \autoref{fig:vc-fwhm}. For the \citet{pascucci2020} sample, stellar masses from \citet{banzatti2019} and \citet{fang2018} are used to normalize the \Neline line widths. Objects without available mass estimates are omitted.
	}
	\label{fig:fwhm-inc}
\end{figure}

\autoref{fig:fwhm-inc} shows the width of the components as a function of disc inclination. The widths are normalized by the square root of the stellar mass to facilitate comparison with the broadening expected from Keplerian rotation.
At inclinations above $i \gtrsim 40\degree$, the component widths fall well within the observed range and are generally consistent with the expected trend from Keplerian broadening across all lines and models, with the exception of the \htline, which deviates from this trend at $i \gtrsim 70\degree$.
This deviation is caused by absorption features that reduce the accuracy of the Gaussian fits.

At lower inclinations, the synthetic profiles become broader than expected from purely Keplerian motion, indicating that thermal broadening or the local velocity gradient begins to dominate.
In the \oiline, the \bsix and \pe models follow the Keplerian trend down to $i \sim 20\degree$, whereas the \bfour model begins to deviate at higher inclinations due to its stronger velocity gradient in the emitting region.
For the other lines, this difference between models is less pronounced, as the emission originates farther out, where the velocity gradients are more similar.
Nevertheless, all models deviate from the Keplerian expectation below $i \lesssim 30\degree$.

These results highlight the need for caution when estimating the emission radius of low-velocity components based on line width alone, particularly for systems with disc inclinations below $i \sim 40\degree$.

\section{Discussion}
\subsection{Can line profiles help to distinguish magnetically from thermally driven winds?}

Spectral line profiles remain one of the few observational tools available to constrain the launching mechanisms of protoplanetary disc winds. However, our results show that certain profile-based diagnostics, such as the inclination at which the centroid velocity peaks, are not robust indicators. Differences in the spatial extent and opening angles of the wind lead to small shifts in peak inclination across models, but these are difficult to constrain observationally, even with high-resolution spectroscopy. This limits their utility as clear discriminants between thermal and magnetic driving.

Instead, the most reliable signatures remain the overall blueshift and FWHM of the low-velocity components.
Strongly magnetised winds, such as in our \bfour model, consistently produce more blueshifted and broader profiles.
This arises from a combination of steeper velocity gradients and emission originating closer to the star.
In contrast, the weakly magnetised or thermally driven models (\bsix and \pe) predict narrower lines with smaller blueshifts, in better agreement with the bulk of the observed narrow low velocity components (NLVCs). This reinforces previous conclusions that most observed NLVCs are more consistent with low magnetisation levels or predominantly photoevaporative winds \citep[e.g.][]{pascucci2020,weber2020,rab2022}.

\subsection{Limitations from the inner radial boundary}

A major challenge for detailed global magnetothermal wind models is the rapidly increasing computational demand as the inner radial boundary is moved closer to the star.
In our simulations, this boundary is set to 0.45\,au, and for the post-processing, we exclude the region inside 0.6\,au to avoid artifacts near the boundary.
While this is already closer to the star than in previous magnetothermal wind models \citep[e.g.][]{wang2019, rodenkirch2019, gressel2020, sarafidou2024}, it still does not capture the innermost disc regions, where compact winds are expected to be launched.
These compact winds are typically associated with the broad low-velocity components (BLVCs) observed in many systems.
In observed spectra, BLVCs often overlap with the NLVCs, making the decomposition into Gaussian components non-trivial. 
Simultaneously fitting both components can bias the inferred centroid and FWHM of the NLVC, particularly when the line profile is significantly broadened by velocity gradients rather than purely Keplerian motion. 
As a result, our synthetic NLVCs may not be directly comparable to observationally derived NLVCs that are affected by this overlap.
One way to mitigate this issue in observations is to constrain the BLVC fit to the red wing of the profile.
Since the red wing is generally less affected by wind-driven asymmetries and velocity gradients than the blue wing, it offers a cleaner diagnostic of Keplerian broadening and may improve the separation of the BLVC from the NLVC emission.

The importance of including the innermost disc regions was also highlighted by \citet{weber2020}, who produced synthetic spectra of a purely photoevaporative and a semi-analytic MHD wind, and in a simple experiment combined the two to mimic a magnetothermal outflow.
While their results suggested that such models combining MHD with photoevaporative winds could reproduce observations with complex, multi-component line profiles, a direct comparison is difficult: our models do not include the innermost disc regions, whereas theirs neither evolved the wind components self-consistently nor accounted for the interaction of an inner wind with stellar EUV and X-ray radiation. Our simulations show that this interaction can be critical for the resulting emission diagnostics.

In particular, our \bfour model illustrates that a dense, magnetically launched inner wind can substantially attenuate high-energy radiation: EUV photons are almost completely absorbed, and a significant portion of the soft X-ray flux is removed before it reaches the outer wind layers.
This radiative screening is a key factor in shaping the observable line emission and in regulating the efficiency of thermal wind launching, with direct implications for disc evolution \citep[see e.g.][]{weder_population_2023}. 
In our current setup, significant attenuation occurs only in the strongly magnetised case, but with a closer inner boundary an additional innermost wind could further enhance the effect, potentially enough that moderately magnetised winds might exhibit similar effects.
A complete understanding will therefore require models that self-consistently couple the wind launching from the innermost disc with its radiative feedback.
Until then, interpretations of wind-tracing emission lines must take into account that stellar radiation may already be significantly filtered by the wind itself.

\section{Conclusions}

We have presented a comparative study of synthetic spectral diagnostics for magnetothermal and photoevaporative disc winds, focusing on three representative models: a relatively strongly magnetised wind ($\beta4$), a more weakly magnetised wind ($\beta6$), and a purely thermal photoevaporative wind (\pe). By post-processing these models with radiative transfer codes, we computed spatially resolved line emissivities and synthetic line profiles for \oiline, \Neline, and \htline, and compared them to recent observational data.

Our analysis shows that the more strongly magnetised $\beta4$ wind produces broader and more blueshifted line profiles across all tracers, particularly at low to intermediate inclinations. This behaviour is consistent with the presence of steeper velocity gradients and more compact, high-velocity emission regions. In contrast, the more weakly magnetised $\beta6$, and the non-magnetised \pe models both yield narrower profiles with smaller blueshifts. These naturally agree better with the majority of observed narrow low-velocity components (NLVCs).

Mapping the spatial distribution of the emission reveals that higher magnetisation leads to increased inner wind column densities, which significantly attenuate EUV and soft X-ray photons. This results in a suppression of [OI] and [NeII] line luminosities, despite higher wind speeds. Interestingly, the \htline emission region in the $\beta4$ model remains similar to the \pe model, suggesting that the effects of attenuation, on the one hand, and increased EUV luminosity, on the other hand, may compensate each other in certain cases.

We find that some line profile diagnostics, such as the inclination at which the centroid velocity peaks, are not robust enough to distinguish the considered wind-driving mechanisms in practice. However, the overall blueshift and FWHM of the low-velocity components remain reliable indicators, with the $\beta4$ model representing an extreme case, that encompasses the most blueshifted observed NLVCs.

Our results thus confirm that relatively strongly magnetised inner winds can imprint distinct spectral signatures -- although these appear to be rare among current observational samples. We conclude that most observed NLVCs are consistent with low magnetisation levels or purely thermal (photoevaporative) winds.

\begin{acknowledgements}
     We are grateful to the anonymous referee for a constructive report which helped improve the manuscript. We acknowledge the support of the Deutsche Forschungsgemeinschaft (DFG, German Research Foundation) Research Unit ``Transition discs'' - 325594231. This research was supported by the Excellence Cluster ORIGINS which is funded by the Deutsche Forschungsgemeinschaft (DFG, German Research Foundation) under Germany's Excellence Strategy - EXC-2094 - 390783311. CHR is grateful for support from the Max Planck Society. OG acknowledges that this work was co-funded\,\footnote{Views and opinions expressed are however those of the author(s) only and do not necessarily reflect those of the European Union or the European Research Council. Neither the European Union nor the granting authority can be held responsible for them.} by the European Union (ERC-CoG, \textsc{Epoch-of-Taurus}, No. 101043302). This research has made use of NASA's Astrophysics Data System. This research made use of Astropy, a community-developed core Python package for Astronomy \citep{AstropyCollaboration2013,AstropyCollaboration2018}, matplotlib \citep{Hunter2007}, numpy \citep{2020NumPy-Array}, and scipy \citep{2020SciPy-NMeth}.
\end{acknowledgements}

\bibliographystyle{bibtex/aa}
\bibliography{bibtex/references_short.bib}

\end{document}